            \documentstyle[12pt,a4]{article}
            \newcommand{\be}{\begin{equation}}
            \newcommand{\ee}{\end{equation}}
\begin{document}
            \vspace{20.0mm}
            \title{Speed Selection Mechanism for Propagating
            Fronts in Reaction-Diffusion Systems with
            Multiple Fields}
            
            \author{{ Stavros Theodorakis$^{a}$ and
            Epameinondas Leontidis$^{b}$}\\
            $^{a}$ {\normalsize \it Physics Department,
            University of Cyprus, P.O. Box 20537,
            Nicosia 1678, Cyprus}\\
            $^b$ {\normalsize \it Chemistry Department,
            University of Cyprus, P.O. Box 20537,
            Nicosia 1678, Cyprus}}

            \maketitle
            
            \vspace{20.0mm}
            \begin{abstract}
            
            \noindent
            We introduce a speed selection mechanism for
            front propagation in reaction-diffusion systems
            with multiple fields. This mechanism applies
            to pulled and pushed fronts alike, and operates
            by restricting the fields to large $finite$
            intervals in the comoving frames of reference.
            The unique velocity for which the center
            of a monotonic solution for a particular field
            is insensitive to the location of the ends
            of the finite interval is the velocity that
            is physically selected for that field,
            making thus the solution approximately
            translation invariant. The fronts for
            the various fields may propagate
            at different speeds, all of them being
            determined though through this mechanism. We
            present analytic
            results for the case of piecewise parabolic
            potentials, and numerical results for
            other cases.
            
            \end{abstract}
            
            \vspace*{5mm}
            \begin{center}
            \end{center}
            \newpage
            \section{The selection mechanism}
            
            In many systems rendered suddenly unstable, propagating
            fronts appear. The determination of the speed of a
            front propagating into an unstable state has attracted
            a lot of attention, since it cannot be achieved
            by simply solving an ordinary differential equation
            in the comoving frame of reference on a one-dimensional
            infinite domain. Indeed, there are many such solutions
            on such a domain, even though the propagating front
            in practice always relaxes to a unique shape and speed.
            The selection principles that have been
            formulated to determine the observable front, without
            having to solve directly the partial differential
            equation of motion for a range of initial conditions,
            have involved concepts of linear and nonlinear
            marginal stability, of structural stability and of
            causality[1], and all of them try to deal with the
            puzzle of the reduction of the multiple solutions to
            the unique observed one. All of these selection
            principles examine the waves from the viewpoint of
            the moving front, the corresponding wave equations
            being reduced then to ordinary differential equations
            involving the speed $v$ of propagation.
            
            \vspace*{3mm}
            These various approaches can be problematic though
            in the case of multiple fields, because not all
            fields need propagate at the same speed, while
            the reduction of the set of partial differential
            equations to a system of ordinary differential equations
            requires that all fields be functions of the
            same variable $x-vt$. More recently, a complete
            analytical understanding of the propagation mechanism
            and relaxation behavior has emerged for those fronts
            that are "pulled along" by the spreading of
            linear perturbations about the unstable state, the
            so called "pulled" fronts[2]. This understanding
            resulted from a detailed study of the relevant partial
            differential equations and explains fully the behavior
            of pulled fronts. The speed selection mechanism
            for those fronts where linear analysis fails, the
            so called "pushed" fronts, is still however the
            subject of ongoing research for the case of multiple
            fields.
            
            \vspace*{3mm}
            The basic problem is that the ordinary
            differential equations that govern the motion of
            uniformly translating fronts do not seem to be able
            to determine the selected velocities for the various
            fields. Indeed, as we said above, the very existence
            of different propagation speeds for the various fields
            makes the examination of the problem from the
            viewpoint of a particular moving frame of reference
            seem irrelevant.
            
            \vspace*{3mm}
            Let us take, for example, the equations
            \begin{eqnarray}
            \frac{\partial\phi_{1}}{\partial t}&=&
            \frac{\partial^{2}\phi_{1}}{\partial x^{2}}
            +\phi_{1}-\phi_{1}^{3},\nonumber\\
            \frac{\partial\phi_{2}}{\partial t}&=&
            D\frac{\partial^{2}\phi_{2}}{\partial x^{2}}
            +\phi_{2}-\phi_{2}^{3}+K\phi_{1},
            \end{eqnarray}
            where $K$ is positive. The dynamics of $\phi_{1}$
            is always independent of that of $\phi_{2}$, for
            fields propagating into the unstable state
            $\phi_{1}=\phi_{2}=0$. If $D<1$, both fields move
            with speed $v=2$. For $D>1$, the $\phi_{1}$ and
            $\phi_{2}$ fronts propagate with different speeds
            $v_{1}=2$ and $v_{2}=2\sqrt{D}$, respectively[2].
            Clearly, the equations indicate that if $\phi_{1}$ is
            a function of $x-vt$ then $\phi_{2}$ should be too.
            Both fields should be propagating therefore with
            the same speed, which is the case however only for
            $D<1$. In fact, $\phi_{2}$ always seems to be moving
            at the maximum available speed. 
            
            \vspace*{3mm}
            It would appear thus that the fronts in
            reaction-diffusion systems with multiple components
            cannot be properly understood in terms of the
            properties of the ordinary differential equations
            that describe uniformly translating solutions. On
            the other hand, the examination of the full
            coupled partial differential equations is
            a rather complicated affair, and there is no
            universal way for dealing with pushed fronts.  
            
            \vspace*{3mm}
            In this paper we present a selection mechanism
            that applies to fronts invading both unstable
            and metastable states, whether they be pulled or
            pushed, and that works even for fields propagating
            at different speeds. Furthermore, it is easy to
            apply, since it involves examination of the system
            from the viewpoint of a single moving frame, resulting
            thus in coupled ordinary differential equations.
            
            \vspace*{3mm}
            This mechanism is the straightforward generalization
            of the speed selection principle presented earlier
            for the case of a single field[3]. It exploits the
            fact that the observed front of a particular field
            is translationally invariant in the comoving frame
            of reference, even on a large $finite$ interval, in
            the sense that its location is effectively
            independent of the ends of the interval. We shall
            be solving then the steady state equations of motion
            on a large finite interval with respect to a
            reference frame moving at an arbitrary given speed
            $v$, subject to the appropriate boundary conditions,
            obtaining a certain solution for each field. The
            solution for a particular field, however, will have
            approximate translational symmetry, thus becoming
            a physically observable front, only for a certain
            value of $v$. It is this value $v^{*}$ of $v$ that
            is experimentally observed during the propagation
            of that field. Thus $the$ $selected$ $front$ $is$
            $the$ $one$ $that$ $is$ $effectively$
            $translationally$ $invariant$ $on$ $a$ $large$
            $finite$ $interval$, in the comoving frame of
            reference. Of course, this selected speed $v^{*}$ will
            not be appropriate, in general, for the other fields,
            in the sense that the corresponding solutions for the
            other fields need not have approximate
            translational symmetry at that speed.
            
            \vspace*{3mm}
            For values of $v$ different from $v^{*}$ the midpoint
            of the field will be either at the left or the right
            end of the finite interval. It is only at $v^{*}$
            that the midpoint can be anywhere near the center of
            the interval, becoming in fact indeterminate. Thus
            a graph of the speed $v$ versus the midpoint of
            the field will have a $plateau$ when $v$ takes the
            value $v^{*}$. The other fields will have such
            plateaus for other values of $v$. It is these plateaus,
            obtained from the graph of $v$ versus the midpoints
            of the fields, that determine the physically
            selected speeds. 
            
            \vspace*{3mm}
            Let us illustrate our mechanism with an example.
            We consider the following reaction-diffusion system:
            \be
            \frac{\partial\phi_{i}}{\partial t}=
            \frac{\partial^{2}\phi_{i}}{\partial x^{2}}
            -\frac{\partial U_{i}(\phi_{i})}{\partial\phi_{i}}
            +\sum_{j\not= i}a_{ij}\phi_{j},
            \ee
            where each $U_{i}$ is a function of the
            corresponding $\phi_{i}$ only. The fixed points of
            this system of differential equations provide
            the appropriate boundary conditions. Let us now
            assume that all fields have monotonic traveling
            wave solutions $\phi_{i}(\xi)$, where $\xi=x-vt$ is
            the coordinate in a given moving frame of
            reference, with $v>0$. Clearly, not all of
            these solutions need have translational symmetry. The
            above partial differential equations reduce then to
            the "steady state" ordinary differential equations
            \be
            \frac{d^{2}\phi_{i}}{d\xi^{2}}
            +v\frac{d\phi_{i}}{d\xi}-\frac{\partial U_{i}(\phi_{i})}
            {\partial\phi_{i}}+\sum_{j\not= i}a_{ij}\phi_{j}=0.
            \ee
            We solve these equations on a large $finite$
            interval [$L_{1}$,$L_{2}$], with $L_{1}\ll L_{2}$,
            subject to the boundary conditions
            $\phi_{j}(L_{1})=p_{j}$ and $\phi_{j}(L_{2})=q_{j}$,
            say, where j runs over all the fields, and where
            $p_{j}$, $q_{j}$ and $a_{ij}$ are constant.
            
            \vspace*{3mm}
            Let us now concentrate on a particular field
            $\phi_{i}$, and let us find the selected velocity of
            the corresponding front. Suppose that $\phi_{i}(\xi)$
            is the solution of Eq. (3) subject to the boundary
            conditions mentioned above. There is only one such
            solution for a given velocity $v$. We multiply now
            Eq. (3) with $d\phi_{i}/d\xi$ and integrate from
            $L_{1}$ to $L_{2}$, obtaining thus
            \be
            v=\frac{U_{i}(q_{i})-U_{i}(p_{i})
            -\frac{1}{2}w_{i}^{2}(L_{2})
            +\frac{1}{2}w_{i}^{2}(L_{1})
            -\sum_{j\not= i}a_{ij}
            \int_{L_{1}}^{L_{2}}w_{i}(\xi)\phi_{j}(\xi)d\xi}
            {\int_{L_{1}}^{L_{2}}w_{i}^{2}(\xi)d\xi},
            \ee
            with $w_{i}(\xi)=d\phi_{i}/d\xi$. If $\phi_{i}(\xi)$
            is going to be a physically observable front on this
            large, but finite, interval, it will have to be
            essentially translationally invariant. This means
            that $d\phi_{i}/d\xi$ will be effectively zero in the
            regions close to the boundaries, $\phi_{i}$ having
            reached its fixed points there. Consequently
            $w_{i}(L_{1})$ and $w_{i}(L_{2})$ will tend to
            zero, while the integrals
            $\int_{L_{1}}^{L_{2}}w_{i}^{2}(\xi)d\xi$
            and $\int_{L_{1}}^{L_{2}}w_{i}(\xi)\phi_{j}(\xi)d\xi$
            will be finite and practically independent of
            $L_{1}$ and $L_{2}$, as $L_{1}\rightarrow -\infty$
            and $L_{2}\rightarrow\infty$. Hence the speed $v$ of
            Eq. (4) becomes independent of the endpoints of
            the interval, acquiring a unique value $v^{*}_{i}$.
            In other words, only the front with that particular
            speed $v^{*}_{i}$ can correspond to an
            essentially translationally invariant field $\phi_{i}$.
            The other fields will not, in general, have
            translational invariance at that particular value
            of $v$, but that does not affect the above argument.
            Indeed, these other fields always appear multiplied
            by the quantity $w_{i}(\xi)$, which is zero in the
            regions near the boundaries when $v$ takes the value
            $v^{*}_{i}$ corresponding to a translationally
            invariant $\phi_{i}(\xi)$. Thus the integral
            $\int_{L_{1}}^{L_{2}}w_{i}(\xi)\phi_{j}(\xi)d\xi$
            remains independent of $L_{1}$ and $L_{2}$, even
            if these other fields have no translational symmetry.
            
            \vspace*{3mm}
            The requirement that the front $\phi_{i}$ be
            independent of the ends of the finite interval,
            when $v=v^{*}_{i}$, selects therefore the speed
            \be
            v^{*}_{i}=\frac{U_{i}(q_{i})-U_{i}(p_{i})
            -\sum_{j\not= i}a_{ij}
            \int_{L_{1}}^{L_{2}}w_{i}(\xi)\phi_{j}(\xi)d\xi}
            {\int_{L_{1}}^{L_{2}}w_{i}^{2}(\xi)d\xi},
            \ee
            with $L_{1}\rightarrow -\infty$ and
            $L_{2}\rightarrow \infty$,
            as the speed of the physically observed front for
            the field $\phi_{i}$. Note that no distinction has
            been made here between metastable and unstable
            states. Indeed, given $any$ particular speed $v$, we
            can find a front interpolating between the stable
            and the unstable or metastable state, provided
            the solution is found on a finite interval. As
            the boundaries go to infinity, the value of the speed
            is restricted to $v^{*}_{i}$ and the front becomes
            the one corresponding to the speed of Eq. (5). 
            
            \section{Analytic example}
            We shall demonstrate the proposed selection
            principle through analytic and numerical work. We
            shall adopt for our analytic work the following
            system of dimensionless partial differential equations:
            \begin{eqnarray}
            \frac{\partial\phi_{1}}{\partial t}&=&
            \frac{\partial^{2}\phi_{1}}{\partial x^{2}}
            +f_{\mu}(\phi_{1}),\nonumber\\
            \frac{\partial\phi_{2}}{\partial t}&=&
            \frac{\partial^{2}\phi_{2}}{\partial x^{2}}
            +f_{\nu}(\phi_{2})+g|\phi_{1}|,
            \end{eqnarray}
            where $g>0$, $\mu\geq 3$, $\nu\geq 3$, and
            \begin{eqnarray}
            f_{\mu}(u)&=&|u|\,\,\,\,\,\,\,\,\,\,\,\,\,\,\,\,\,
            if\,\,\,\,\,|u|\leq 1/2,\nonumber\\
            &=&\mu(1-|u|)\,\,\,\,\,if\,\,\,\,\,|u|\geq 1/2.
            \end{eqnarray}
            This piecewise linear choice for the function
            $f_{\mu}(u)$ results from a piecewise parabolic
            potential and will lead to exact analytic solutions.
            Piecewise linear representations of nonlinearities
            have often provided an analytically rigorous basis
            for the study of diffusion systems[4], as well as
            of nucleation and crystallization problems[5], always
            on the interval $(-\infty,\infty)$.
            
            \vspace*{3mm}
            We shall be looking for uniformly translating
            solutions, functions of the variable $\xi=x-vt$,
            where $v>0$ is an arbitrary given speed. Thus Eqs.
            (6) become
            \be
            \frac{d^{2}\phi_{1}}{d\xi^{2}}
            +v\frac{d\phi_{1}}{d\xi}
            +f_{\mu}(\phi_{1})= 0,
            \ee
            and
            \be
            \frac{d^{2}\phi_{2}}{d\xi^{2}}
            +v\frac{d\phi_{2}}{d\xi}
            +f_{\nu}(\phi_{2})+g|\phi_{1}|=0.
            \ee
            
            \vspace*{3mm}
            The mirror symmetries of Eqs. (8) and (9)
            allow us to work with positive fields only. So we
            shall assume, without loss of generality, that
            $\phi_{1}\geq 0$, $\phi_{2}\geq 0$. When the fields
            are constant in
            space and time, they are at their fixed points. These
            fixed points ($\phi_{1}$, $\phi_{2}$) are the
            points (0,0) (unstable fixed point), (0,1)
            (saddle point) and (1,1+$g/\nu$)
            (stable fixed point). We shall
            be interested in fronts invading the unstable
            state $\phi_{1}=0$, $\phi_{2}=0$, so we need to
            solve Eqs. (8) and (9) on the $finite$ interval
            [$L_{1}$, $L_{2}$], subject to the boundary conditions
            $\phi_{1}(L_{1})=1$, $\phi_{2}(L_{1})=1+g/\nu$,
            $\phi_{1}(L_{2})=0$, $\phi_{2}(L_{2})=0$, where
            $L_{1}\ll 0\ll L_{2}$. These boundary conditions
            ensure that the system makes a phase transition from
            the unstable state to the stable state. Furthermore,
            we shall define the midpoints $\xi_{1}$ and $\xi_{2}$
            of the fields $\phi_{1}$ and $\phi_{2}$ through
            the relations $\phi_{1}(\xi_{1})=1/2$ and
            $\phi_{2}(\xi_{2})=1/2$, respectively, noting that
            the fields and their slopes have to be continuous
            at these points.
            
            \vspace*{3mm}
            The dynamics of $\phi_{1}$ is decoupled from
            the dynamics of $\phi_{2}$, consequently we can
            easily find the corresponding solution. There
            are five boundary conditions that must be
            satisfied, namely two at the edges, the
            continuity of $\phi_{1}$ and
            of $d\phi_{1}/d\xi$ at $\xi_{1}$, and the definition
            of $\xi_{1}$. On the other hand, the solution of
            Eq. (8) for the field $\phi_{1}$ will involve
            five unknown constants for any given value of
            $v$, namely $\xi_{1}$ and two constants for each
            linear piece of $f_{\mu}(\phi_{1})$. We expect
            therefore a unique solution $\phi_{1}(\xi)$ for
            each value of $v$.  
            
            \vspace*{3mm}
            Indeed, the exact solution of Eq. (8) for the field
            $\phi_{1}$ is 
            \begin{eqnarray}
            \phi_{1}(\xi)&=&1-\frac{e^{m_{1}(\xi-L_{1})}
            -e^{m_{2}(\xi-L_{1})}}{2e^{m_{1}(\xi_{1}-L_{1})}
            -2e^{m_{2}(\xi_{1}-L_{1})}}\,\,\,\,\,if\,\,\,\,
            L_{1}\leq\xi\leq\xi_{1}\nonumber\\
            &=&\frac{e^{k_{1}(\xi-L_{2})}-
            e^{k_{2}(\xi-L_{2})}}{2e^{k_{1}(\xi_{1}-L_{2})}
            -2e^{k_{2}(\xi_{1}-L_{2})}}\,\,\,\,\,
            if\,\,\,\,\xi_{1}\leq\xi\leq L_{2},
            \end{eqnarray}
            where
            \be
            k_{1}=\frac{1}{2}(-v+\sqrt{v^{2}-4}),
            \ee
            \be
            k_{2}=\frac{1}{2}(-v-\sqrt{v^{2}-4}),
            \ee
            \be
            m_{1}=\frac{1}{2}(-v+\sqrt{v^{2}+4\mu}),
            \ee
            \be
            m_{2}=\frac{1}{2}(-v-\sqrt{v^{2}+4\mu}),
            \ee
            and $\xi_{1}$ satisfies
            \be
            \frac{k_{1}e^{k_{1}(\xi_{1}-L_{2})}
            -k_{2}e^{k_{2}(\xi_{1}-L_{2})}}{e^{k_{1}
            (\xi_{1}-L_{2})}-e^{k_{2}(\xi_{1}-L_{2})}}
            =-\frac{m_{1}e^{m_{1}(\xi_{1}-L_{1})}
            -m_{2}e^{m_{2}(\xi_{1}-L_{1})}}{e^{m_{1}
            (\xi_{1}-L_{1})}-e^{m_{2}(\xi_{1}-L_{1})}}.
            \ee
            The solution of Eq. (15) gives $\xi_{1}$ as a
            function of the speed $v$. We note that $m_{1}>0>m_{2}$
            and $k_{2}<k_{1}<0$. For a given value of $v$, Eqs.
            (10)-(15) determine fully the field $\phi_{1}(\xi)$.
            
            \vspace*{3mm}
            As shown in our earlier work[3], the graph of $v$
            versus $\xi_{1}$ has a plateau at $v=v_{c1}$,
            where $v_{c1}=$$(\mu+1)/\sqrt{2\mu-2}$. Indeed, if
            we require $L_{1}\ll\xi_{1}\ll L_{2}$, then
            Eq. (15) reduces to $k_{2}+m_{1}=0$. This equation has
            a real solution, $v=v_{c1}$, provided $\mu\geq 3$.
            In other words, for that particular value of $v$
            the midpoint $\xi_{1}$ can be anywhere in the interval
            and cannot be determined, rendering thus the
            front effectively translationally invariant on the
            finite domain $[L_{1},L_{2}]$. The value $v_{c1}$ is
            therefore the selected speed of $\phi_{1}$ if
            $\mu\geq 3$ (pushed case). If $v>v_{c1}$, then
            $\xi_{1}$ is close to $L_{1}$, while for $v<v_{c1}$
            we find that $\xi_{1}$ lies close to $L_{2}$.
            The existence of the plateau at $v_{c1}$ shows
            thus that the only value of $v$ for which the
            field $\phi_{1}$ is approximately translationally
            invariant is $v_{c1}$.
            
            \vspace*{3mm}
            Let us now turn our attention to the field
            $\phi_{2}$. The relevant equation must be solved on
            the three segments determined by $L_{1}$, $L_{2}$,
            $\xi_{1}$ and $\xi_{2}$. There are seven boundary
            conditions: two at the edges, four from the
            continuity of $\phi_{2}$ and $d\phi_{2}/d\xi$ at
            the points $\xi_{1}$ and $\xi_{2}$, and the
            definition of $\xi_{2}$. There are also seven
            unknown quantities for a given value of $v$: two on
            each of the three segments, and $\xi_{2}$. Note
            that Eq. (15) determines the other midpoint. We
            expect therefore a unique solution $\phi_{2}(\xi)$
            for each given value of $v$.
            
            \vspace*{3mm}
            Clearly, if $g$ were zero, then the speed of
            propagation for $\phi_{2}$ would be
            $v_{c2}=(\nu+1)/\sqrt{2\nu-2}$, and the two fields would
            be completely decoupled. We need to examine what
            happens for $g\not= 0$. We can obtain some
            qualitative results by looking at Eqs. (8) and (9).
            We distinguish two cases.
            
            \vspace*{3mm}
            ${\bf (a)}$ ${\bf Case}$ ${\bf v_{c2}>v_{c1}}$
            
            \vspace*{3mm}
            (i) We examine the case $v>v_{c2}>v_{c1}$ first.
            Since $v>v_{c1}$, we have $\xi_{1}\approx L_{1}$,
            as discussed in our earlier work[3]. Therefore the
            field $\phi_{1}$ falls very rapidly from 1 to 0, and
            it remains equal to 0 on almost all of the
            interval [$L_{1},L_{2}$]. Since $\phi_{1}$ is
            approximately zero almost everywhere, Eqs. (8) and (9)
            decouple. Thus $\phi_{2}$ behaves as if $g$ were
            equal to 0. Since $v>v_{c2}$, this implies that
            $\xi_{2}\approx L_{1}$. Hence $\phi_{2}$ is also
            zero practically everywhere. Both fields are
            essentially on the fixed point (0,0).
            
            \vspace*{3mm}
            (ii) We examine the case $v_{c2}>v>v_{c1}$ next.
            Again $\xi_{1}\approx L_{1}$, since $v>v_{c1}$,
            hence $\phi_{1}$ is essentially zero on almost all
            of the interval. The fields decouple once more, but
            now $\xi_{2}\approx L_{2}$, since $v<v_{c2}$. Hence
            the field $\phi_{2}$ is nonzero up to the point
            $\xi=L_{2}$. The only fixed point that is available
            for the two fields is then the point (0,1).
            Thus $\phi_{1}$ starts out at $L_{1}$ having the value
            1 and very rapidly drops down to zero. The
            field $\phi_{2}$, on the other hand, starts out
            having the value $1+g/\nu$ at $L_{1}$, drops down to
            the value 1 almost immediately, and then it stays
            there till it reaches the other edge, where it
            drops down to zero. Thus the midpoint of
            $\phi_{2}$ shifts abruptly from the left edge to
            the right edge the very moment we pass from case (i) to
            case (ii), i.e. at $v=v_{c2}$, because $\xi_{2}
            \approx L_{1}$ for $v>v_{c2}$, but $\xi_{2}\approx
            L_{2}$ for $v<v_{c2}$.   
            
            \vspace*{3mm}
            (iii) We examine finally the case $v_{c2}>v_{c1}>v$.
            Since $v_{c1}>v$, we have $\xi_{1}\approx L_{2}$.
            Therefore $\phi_{1}$ remains on the value 1 on almost
            all of the interval, dropping down to zero only very
            close to the right edge. The only available fixed
            point for the two fields is then the
            point $(1,1+g/\nu)$. Therefore the field $\phi_{2}$
            remains stuck at the value $1+g/\nu$ almost everywhere,
            dropping to zero only very close to the right edge,
            whereby $\xi_{2}\approx L_{2}$.
            We note that the abrupt shift of the midpoint of
            $\phi_{1}$ occurs at $v=v_{c1}$.
            
            \vspace*{3mm}
            These arguments indicate then that the field
            $\phi_{1}$ acquires approximate translational
            invariance when $v=v_{c1}$, as expected, since
            its dynamics is decoupled from the dynamics
            of $\phi_{2}$. At that speed we have a plateau of
            $v$ versus $\xi_{1}$. On the other hand, the plateau
            of $v$ versus $\xi_{2}$ occurs at $v=v_{c2}$. Therefore
            the front of $\phi_{2}$ propagates with the speed
            $v_{c2}$. Indeed then, our finite interval mechanism
            gives the selected velocities for both fields, in
            spite of their being different. We also note that in
            all likelihood $\xi_{1}<\xi_{2}$, since in the region
            $v_{c1}<v<v_{c2}$ we found that $\xi_{1}\approx L_{1}$
            and $\xi_{2}\approx L_{2}$.
            
            \vspace*{3mm}
            ${\bf (b)}$ ${\bf Case}$ ${\bf v_{c2}<v_{c1}}$
            
            \vspace*{3mm}
            (i) We examine the case $v>v_{c1}>v_{c2}$ first. Since
            $v>v_{c1}$, we have $\xi_{1}\approx L_{1}$, as
            discussed in our earlier work[3]. Therefore the
            field $\phi_{1}$ falls very rapidly from 1 to 0, and
            it remains equal to 0 on almost all of the interval
            [$L_{1},L_{2}$]. Since $\phi_{1}$ is approximately zero
            almost everywhere, Eqs. (8) and (9) decouple. Thus
            $\phi_{2}$ behaves as if $g$ were equal to 0.
            Since $v>v_{c2}$, this implies that
            $\xi_{2}\approx L_{1}$. Hence $\phi_{2}$ is also
            zero practically everywhere. Both fields are
            essentially on the fixed point (0,0).
            
            \vspace*{3mm}
            (ii) We examine the case $v_{c1}>v>v_{c2}$ next.
            Now $\xi_{1}\approx L_{2}$, since $v<v_{c1}$,
            hence $\phi_{1}$ is stuck on the value 1 on almost
            all of the interval. The only fixed point that
            is available for the two fields then is the
            point $(1,1+g/\nu)$. That means that $\phi_{2}$ must
            be stuck at the value $1+g/\nu$ on almost all of
            the interval, dropping down to zero only close to
            the right edge. Hence $\xi_{2}\approx L_{2}$. Thus
            both the midpoints of $\phi_{2}$ and $\phi_{1}$
            shift suddenly from the left edge to the right edge
            the very moment we pass from case (i) to case (ii),
            i.e. at $v=v_{c1}$.   
            
            \vspace*{3mm}
            (iii) We examine finally the case $v_{c1}>v_{c2}>v$.
            Since $v_{c1}>v$, we have $\xi_{1}\approx L_{2}$.
            Therefore $\phi_{1}$ remains on the value 1 on almost
            all of the interval, dropping down to zero only
            very close to the right edge. The only available fixed
            point for the two fields is once more the point
            $(1,1+g/\nu)$. Therefore the field $\phi_{2}$
            remains stuck at the value $1+g/\nu$ almost
            everywhere, dropping to zero only very close to
            the right edge, whereby $\xi_{2}\approx L_{2}$.
            
            \vspace*{3mm}
            These arguments indicate then that the field
            $\phi_{1}$ acquires approximate translational
            invariance when $v=v_{c1}$, since its dynamics is
            decoupled from the dynamics of $\phi_{2}$. At that
            speed we have a plateau of $v$ versus $\xi_{1}$.
            On the other hand, the plateau of $v$ versus
            $\xi_{2}$ occurs also at $v=v_{c1}$. Therefore the
            front of $\phi_{2}$ propagates with the speed
            $v_{c1}$. In this case both fields propagate at
            the same speed.
            
            \vspace*{3mm}
            We note that the field $\phi_{2}$
            always propagates at the maximum possible speed,
            i.e. $v_{c2}$ in case (a) and $v_{c1}$ in case
            (b), just like the fields of Eqs. (1). Our
            finite interval mechanism is able to handle
            both cases though. We also note that the speeds of
            propagation are independent of the coupling constant
            $g$, irrespective of how large or small it is.
            
            \vspace*{3mm}
            Let us now verify this behavior by solving
            analytically Eq. (9) to find $\phi_{2}$, given the
            solution of Eq. (10) for the field $\phi_{1}$.
            
            \vspace*{3mm}
            We shall assume that $\xi_{1}\leq\xi_{2}$,
            for the sake of definiteness.
            This situation is appropriate for the
            case $v_{c2}>v_{c1}$, according to the
            arguments presented above. If contradictions
            arise due to this assumption, it will be easy
            enough to repeat the work with the contrary
            assumption. In fact, it turns out that the
            relation $\xi_{1}\leq\xi_{2}$ holds even if
            $v_{c2}<v_{c1}$, in the examples we shall present.
            
            \vspace*{3mm}
            Let us examine the region $\xi_{2}\leq\xi\leq L_{2}$
            first. The boundary conditions $\phi_{2}(\xi_{2})=1/2$
            and $\phi_{2}(L_{2})=0$ determine the two constants
            that will appear in the solution of the
            ordinary differential equation on this interval.
            Thus the full solution for $\phi_{2}$ on the
            interval $[\xi_{2}, L_{2}]$ turns out to be
            \be
            \phi_{2}(\xi)=(\xi-L_{2})
            \Bigl(-\frac{e^{k_{1}(\xi-L_{2})}}{2k_{1}+v}
            +\frac{e^{k_{2}(\xi-L_{2})}}{2k_{2}+v}\Bigr)z_{1}
            +\gamma(e^{k_{1}(\xi-L_{2})}-e^{k_{2}(\xi-L_{2})}),
            \ee
            where
            \be
            z_{1}=g/(2e^{k_{1}(\xi_{1}-L_{2})}
            -2e^{k_{2}(\xi_{1}-L_{2})}),
            \ee
            and
            \be
            \gamma=\frac{\frac{1}{2}
            -(\xi_{2}-L_{2})
            \Bigl(-e^{k_{1}(\xi_{2}-L_{2})}/(2k_{1}+v)
            +e^{k_{2}(\xi_{2}-L_{2})}/(2k_{2}+v)\Bigr)z_{1}}
            {e^{k_{1}(\xi_{2}-L_{2})}
            -e^{k_{2}(\xi_{2}-L_{2})}}.
            \ee
            Eq. (16) now yields the quantity
            $\phi_{2}^{'}(\xi_{2})$,
            \begin{eqnarray}
            \phi_{2}^{'}(\xi_{2})&=&
            \Bigl(-\frac{e^{k_{1}(\xi_{2}-L_{2})}}{2k_{1}+v}
            +\frac{e^{k_{2}(\xi_{2}-L_{2})}}{2k_{2}+v}\Bigr)z_{1}
            +\frac{2z_{1}(\xi_{2}-L_{2})}
            {e^{-k_{2}(\xi_{2}-L_{2})}
            -e^{-k_{1}(\xi_{2}-L_{2})}}\nonumber\\
            &+&\frac{k_{1}e^{k_{1}(\xi_{2}-L_{2})}
            -k_{2}e^{k_{2}(\xi_{2}-L_{2})}}
            {2e^{k_{1}(\xi_{2}-L_{2})}-2e^{k_{2}(\xi_{2}-L_{2})}}.
            \end{eqnarray}
            We can now use the known values of
            $\phi_{2}(\xi_{2})$ and $\phi_{2}^{'}(\xi_{2})$ as
            boundary conditions in order
            to solve Eq. (9) on the interval $[\xi_{1},\xi_{2}]$.
            We find
            \be
            \phi_{2}(\xi)=1+\frac{z_{1}(e^{k_{1}(\xi-L_{2})}
            -e^{k_{2}(\xi-L_{2})})}{\nu+1}+Ae^{n_{1}(\xi-L_{2})}
            +Be^{n_{2}(\xi-L_{2})},
            \ee
            where
            \be
            n_{1}=\frac{1}{2}(-v+\sqrt{v^{2}+4\nu}),
            \ee
            \be
            n_{2}=\frac{1}{2}(-v-\sqrt{v^{2}+4\nu}),
            \ee
            with $n_{1}>0$ and $n_{2}<0$,
            \be
            \Omega_{1}=-\frac{1}{2}-\frac{z_{1}}{\nu+1}\bigl(e^{k_{1}
            (\xi_{2}-L_{2})}-e^{k_{2}(\xi_{2}-L_{2})}\bigr),
            \ee
            \be
            \Omega_{2}=\phi_{2}^{'}(\xi_{2})
            -\frac{z_{1}}{\nu+1}\bigl(k_{1}e^{k_{1}
            (\xi_{2}-L_{2})}-k_{2}e^{k_{2}(\xi_{2}-L_{2})}\bigr),
            \ee
            \be
            A=e^{-n_{1}(\xi_{2}-L_{2})}
            \frac{\Omega_{2}-n_{2}\Omega_{1}}{n_{1}-n_{2}},
            \ee
            \be
            B=e^{-n_{2}(\xi_{2}-L_{2})}
            \frac{\Omega_{2}-n_{1}\Omega_{1}}{n_{2}-n_{1}}.
            \ee
            We can now use Eq. (20) to find $\phi_{2}(\xi_{1})$ and
            $\phi_{2}^{'}(\xi_{1})$. We get
            \be
            \phi_{2}(\xi_{1})=1+\frac{g}{2+2\nu}+Ae^{n_{1}
            (\xi_{1}-L_{2})}+Be^{n_{2}(\xi_{1}-L_{2})}
            \ee
            and
            \be
            \phi_{2}^{'}(\xi_{1})=
            \bigl(k_{1}e^{k_{1}(\xi_{1}-L_{2})}
            -k_{2}e^{k_{2}(\xi_{1}-L_{2})}\bigr)\frac{z_{1}}{\nu+1}
            +n_{1}Ae^{n_{1}(\xi_{1}-L_{2})}
            +n_{2}Be^{n_{2}(\xi_{1}-L_{2})}.
            \ee
            Finally, we can use these values of $\phi_{2}$ and
            $\phi_{2}^{'}$ at $\xi_{1}$ in order to find the
            solution to $\phi_{2}$ in the
            interval $[L_{1},\xi_{1}]$. Imposing then
            the boundary condition $\phi_{2}(L_{1})=1+g/\nu$
            on this solution leads to the final relation
            \begin{eqnarray}
            &&\Omega_{2}-n_{1}\Omega_{1}=\nonumber\\
            &&\Bigl(\frac{g}{2\mu-2\nu}
            +\frac{g}{2+2\nu}\Bigr)\frac{k_{1}
            e^{k_{1}(\xi_{1}-L_{2})}-k_{2}
            e^{k_{2}(\xi_{1}-L_{2})}}{e^{k_{1}(\xi_{1}-L_{2})}
            -e^{k_{2}(\xi_{1}-L_{2})}}
            \bigl(e^{-(n_{1}-n_{2})(\xi_{1}-L_{1})}-1\bigr)
            e^{n_{2}(\xi_{2}-\xi_{1})}\nonumber\\
            &+&(\Omega_{2}-n_{2}\Omega_{1})e^{(n_{1}-n_{2})
            (-\xi_{2}+L_{1})}\nonumber\\
            &+&\Bigl(\frac{g}{\nu}+\frac{g}{2\mu-2\nu}
            -\frac{g}{2+2\nu}\Bigr)\bigl(n_{2}e^{-(n_{1}-n_{2})
            (\xi_{1}-L_{1})}-n_{1}\bigr)e^{n_{2}(\xi_{2}-\xi_{1})}.
            \end{eqnarray}
            We can use this equation to find the plateau of
            $\phi_{2}$.
            
            \vspace*{3mm}
            Indeed, let us take the case $v_{c2}>v_{c1}$ first.
            If $v>v_{c1}$, then we must have $\xi_{1}\approx
            L_{1}$. In that case we can show that any $\xi_{2}$
            that is far from $L_{1}$ and $L_{2}$ will satisfy Eq.
            (29), provided $k_{2}+n_{1}=0$, a relation equivalent
            to the requirement that $v$ be equal to $v_{c2}$. Thus
            $\phi_{2}$ has a plateau above $v_{c1}$, at the speed
            $v=v_{c2}$. Note also that the midpoint $\xi_{2}$ of
            $\phi_{2}$ is already at $L_{2}$ when the midpoint of
            $\phi_{1}$ shifts to $L_{2}$.
            
            \vspace*{3mm}
            On the other hand, if we look at the case
            $v_{c2}<v_{c1}$, then we see that below $v_{c1}$ we
            must have $\xi_{1}\approx L_{2}$. But since all the
            above analytic equations have been derived under the
            assumption $\xi_{1}\leq\xi_{2}$, we conclude that
            $\xi_{2}$ must be close to $L_{2}$ as well. Thus, when
            the midpoint $\xi_{1}$ of $\phi_{1}$ shifts abruptly
            to the right edge, it forces the midpoint of the other
            field to go there as well, provided the
            analytic equations have solutions consistent with the
            assumption $\xi_{1}<\xi_{2}$. The corresponding
            behavior is illustrated by the examples of Figs. 1 and
            2, verifying thus the qualitative conclusions drawn
            earlier. In particular, these figures confirm the
            assumption $\xi_{1}<\xi_{2}$. 
            
            \vspace*{3mm}
            We see then that, if $v_{c2}>v_{c1}$, the field
            $\phi_{2}$ has a plateau at the highest speed
            $v_{c2}$. If, on the other hand,  $v_{c2}<v_{c1}$,
            then $\phi_{1}$ pulls
            $\phi_{2}$ and forces it to propagate at the higher
            speed $v_{c1}$. This behavior is seen even when the
            coupling constant $g$ takes very small values, and
            matches the behavior of the fields that obey Eq. (1).
            
            \section{Numerical examples}
            We can demonstrate our selection mechanism numerically
            as well. Let us examine the following system:
            \begin{eqnarray}
            \frac{\partial\phi_{1}}{\partial t}&=&
            \frac{\partial^{2}\phi_{1}}{\partial x^{2}}
            +h_{b_{1}}(\phi_{1}),\nonumber\\
            \frac{\partial\phi_{2}}{\partial t}&=&
            \frac{\partial^{2}\phi_{2}}{\partial x^{2}}
            +h_{b_{2}}(\phi_{2})+g\phi_{1},
            \end{eqnarray}
            where
            \be
            h_{b}(u)=\frac{1}{b}u(1-u)(b+u).
            \ee
            It was this particular choice of $h(u)$ that was used
            for the case of a single field
            when the concepts of linear and nonlinear marginal
            stability were first proposed[6]. That study found that
            for $0<b<1/2$ the selected speed of the single field
            for the front invading the unstable state is
            $(2b+1)/\sqrt{2b}$.
            
            \vspace*{3mm}
            We shall consider values of $b$ less than $1/2$ (pushed
            case). Thus, if the coupling constant $g$ were zero,
            then the two fields $\phi_{1}$ and $\phi_{2}$ of
            Eq. (30) would propagate separately, with speeds
            $v_{c1}=(2b_{1}+1)/\sqrt{2b_{1}}$ and
            $v_{c2}=(2b_{2}+1)/\sqrt{2b_{2}}$, respectively.
            
            \vspace*{3mm}
            We have solved Eqs. (30) numerically on a finite $\xi$
            domain for the $h(u)$ of Eq. (31), assuming that
            both fields are functions of the variable $\xi=x-vt$,
            with g=21, $b_{1}=1/8$ and $b_{2}=1/18$, subject to the
            boundary conditions $\phi_{1}(L_{1})=1$,
            $\phi_{1}(L_{2})=0$,
            $\phi_{2}(L_{2})=0$ and $\phi_{2}(L_{1})=1.5$. These
            values correspond to the stable and unstable fixed
            points (1,1.5) and (0,0) for the fields $\phi_{1}$ and
            $\phi_{2}$, the points in other words
            where Eqs. (30) acquire uniform solutions. The solutions
            that interpolate between these two fixed points are
            shown in Fig. 3. We can see that the two fields are
            at the saddle point $\phi_{1}=0$, $\phi_{2}=1$, on
            almost all of the interval. This feature reminds us of
            the dual fronts, where the decomposition from the
            unstable to the stable state proceeds via an
            intermediate saddle point, with the two fields
            propagating at different speeds[7].
            
            \vspace*{3mm}
            Fig. 4 shows the locations $\xi_{1}$ and $\xi_{2}$
            where the fields $\phi_{1}$ and $\phi_{2}$
            attain the value $1/2$, at a given arbitrary speed $v$,
            when $g=21$, $b_{1}=1/8$ and $b_{2}=1/18$.
            We can deduce from this figure that the
            field $\phi_{1}$ propagates with speed $v_{c1}=2.5$,
            while the field $\phi_{2}$ propagates with speed
            $v_{c2}=10/3$. Indeed, we see that at these velocities
            there are plateaus of $v$ versus
            the location where each field acquires the value 1/2. In
            other words, the location of the midpoint of the front
            for the field $\phi_{1}$ or $\phi_{2}$
            is indeterminate at the corresponding velocity $v_{c1}$
            or $v_{c2}$, rendering the solution essentially
            translation invariant there. We note that the two
            fronts propagate at different speeds, and that
            $\xi_{1}<\xi_{2}$.
            
            \vspace*{3mm}
            We have also solved Eqs. (30) for the case $g=21$,
            $b_{1}=1/18$ and $b_{2}=1/8$. The unstable fixed point
            is still the point $\phi_{1}=\phi_{2}=0$, but the stable
            fixed point is the point $\phi_{1}=1$,
            $\phi_{2}=1.7768$, since these values satisfy Eqs. (30).
            Hence the solutions we seek
            have to interpolate between these two fixed points,
            subject to the boundary conditions $\phi_{1}(L_{1})=1$,
            $\phi_{1}(L_{2})=0$, $\phi_{2}(L_{2})=0$ and
            $\phi_{2}(L_{1})=1.7768$.
            Fig. 5 shows the midpoints $\xi_{1}$ and $\xi_{2}$ of
            the two fields at an arbitrary speed $v$. We observe
            once more that there are the usual plateaus, $\xi_{1}$
            being again less than $\xi_{2}$. However,
            both plateaus occur at the speed $v=v_{c1}=10/3$. The
            field $\phi_{2}$ is pulled by $\phi_{1}$ and is forced
            to propagate at $v_{c1}$, rather than at its own lower
            speed $v_{c2}=2.5$. The existence of the common
            plateau indicates once again that at that particular
            speed $v_{c1}$ the locations of the midpoints of the
            fields become indeterminate, making thus the two
            fields effectively translation invariant.
            
            \section{Concluding remarks}
            We see then that requiring the solution of a field to
            have approximate translational invariance
            on a finite interval in the comoving frame of reference
            results in the selection of a speed for the front. We
            can adopt then a selection principle that reads "the
            selected front is the one that is approximately
            translationally invariant on a large finite interval,
            with respect to the comoving frame of reference". This
            principle is very easy to implement, especially
            numerically. Indeed, it suffices to solve the
            moving frame equation on a large finite interval,
            for an arbitrary propagation speed.
            For large speeds we expect the midpoint of the
            front to be close to the left boundary. As the speed
            is lowered, the midpoint suddenly moves to the
            right boundary. The speed $v^{*}$ at which
            this sudden move occurs is the speed selected by the
            physically observed front.
            
            \vspace*{3mm}
            Of course, the selection of a certain plateau for a
            particular field need not include the selection of such
            a plateau for other fields as well. In fact, the
            advantage of our mechanism is that the graph of the
            speed of propagation versus the midpoint of each field
            allows us to find the plateaus for all the fields, and
            hence the corresponding physically selected velocities.
            The approximate translational invariance that each field
            acquires as its plateau is reached need not involve the
            other fields as well.
            
            \vspace*{3mm}
            Thus our mechanism can find the selected velocities for
            all the fields present in a system of coupled partial
            differential equations, by simply solving a much
            simpler system of coupled ordinary differential
            equations, having assumed that all the fields are
            traveling at the same arbitrary speed $v$. The value
            of $v$ for which the midpoint of a particular field
            becomes indeterminate is the physically selected
            velocity for that field.

            \newpage
            %

            \newpage
            \noindent
            {\bf Figure Captions \hfill}
            
            \begin{enumerate}
            
            \item[\bf Figure 1:]
            The speed $v$ as a function of the midpoints
            of the fronts invading the unstable state, for the
            system of Eqs. (8) and (9), with $L_{1}=-10$ and
            $L_{2}=10$, $g=9.5$, $\mu=9$, $\nu=19$.
            The plateau is at $v=2.5$ for $\phi_{1}$, and at
            $v=10/3$ for $\phi_{2}$. All quantities are
            dimensionless.
            
            \item[\bf Figure 2:]
            The speed $v$ as a function of the midpoints
            of the fronts invading the unstable state, for the
            system of Eqs. (8) and (9), with $L_{1}=-10$ and
            $L_{2}=10$, $g=4.5$, $\mu=19$, $\nu=9$.
            The plateau is at $v=10/3$ for both $\phi_{1}$ and
            $\phi_{2}$. All quantities are
            dimensionless.
            
            \item[\bf Figure 3:]
            The solutions of Eqs. (30) that interpolate between
            the fixed points (0,0) and (1,1.5), for $L_{1}=-15$,
            $L_{2}=15$, $v=2.7$, $g=21$, $b_{1}=1/8$, $b_{2}=1/18$.
            These solutions lie on the saddle point (0,1) on most
            of the interval.
            
            \item[\bf Figure 4:]
            The speed $v$ as a function of the midpoints
            of the fronts invading the unstable state, for the
            system of Eqs. (30), with $L_{1}=-15$ and
            $L_{2}=15$, $g=21$, $b_{1}=1/8$, $b_{2}=1/18$.
            The plateau is at $v=2.5$ for $\phi_{1}$ and at
            $v=10/3$ for $\phi_{2}$. All quantities are
            dimensionless.
            
            \item[\bf Figure 5:]
            The speed $v$ as a function of the midpoints
            of the fronts invading the unstable state, for the
            system of Eqs. (30), with $L_{1}=-15$ and
            $L_{2}=15$, $g=21$, $b_{1}=1/18$, $b_{2}=1/8$.
            The plateau is at $v=10/3$ for both $\phi_{1}$
            and $\phi_{2}$. All quantities are
            dimensionless.

            \end{enumerate}
            
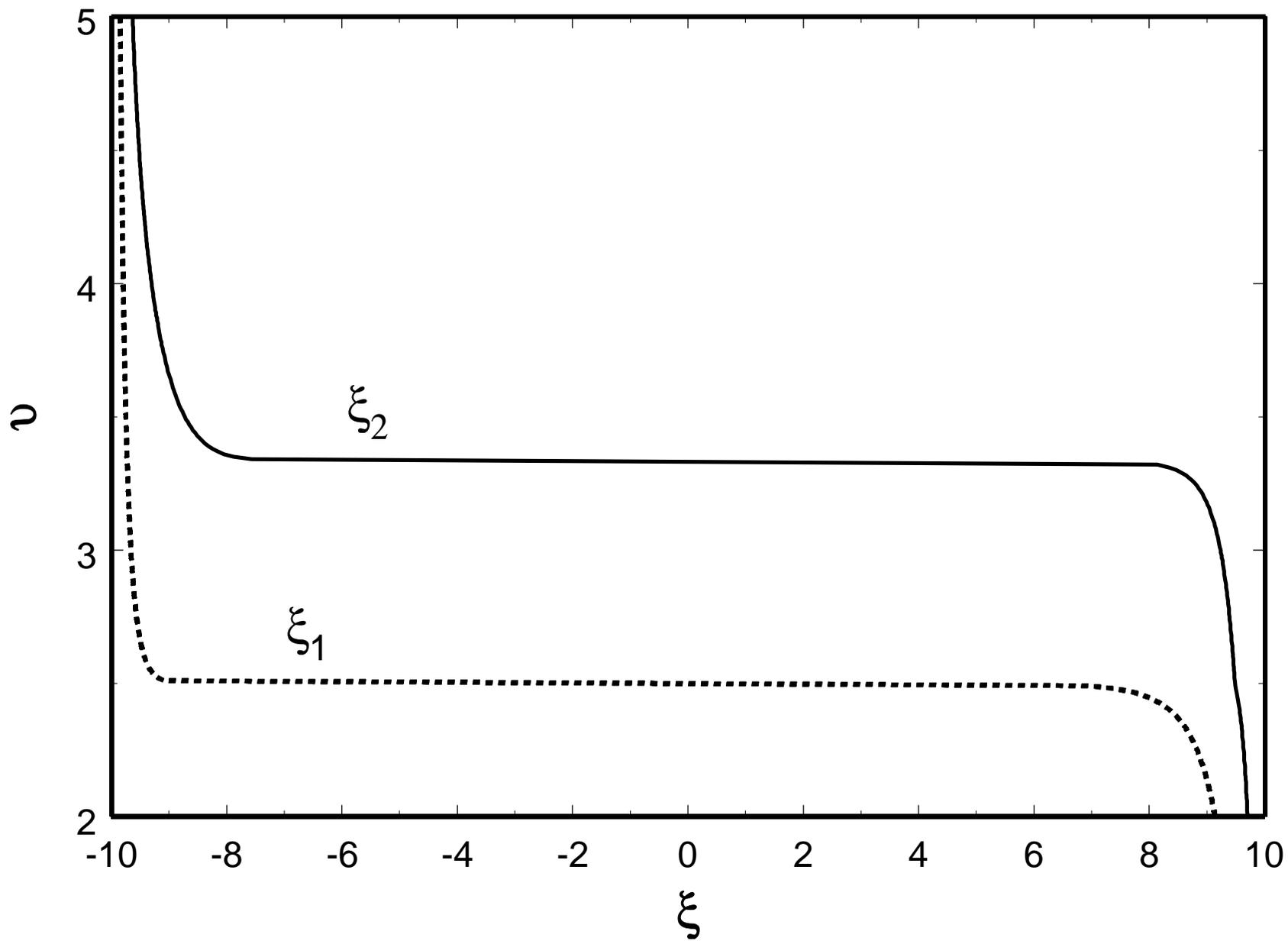
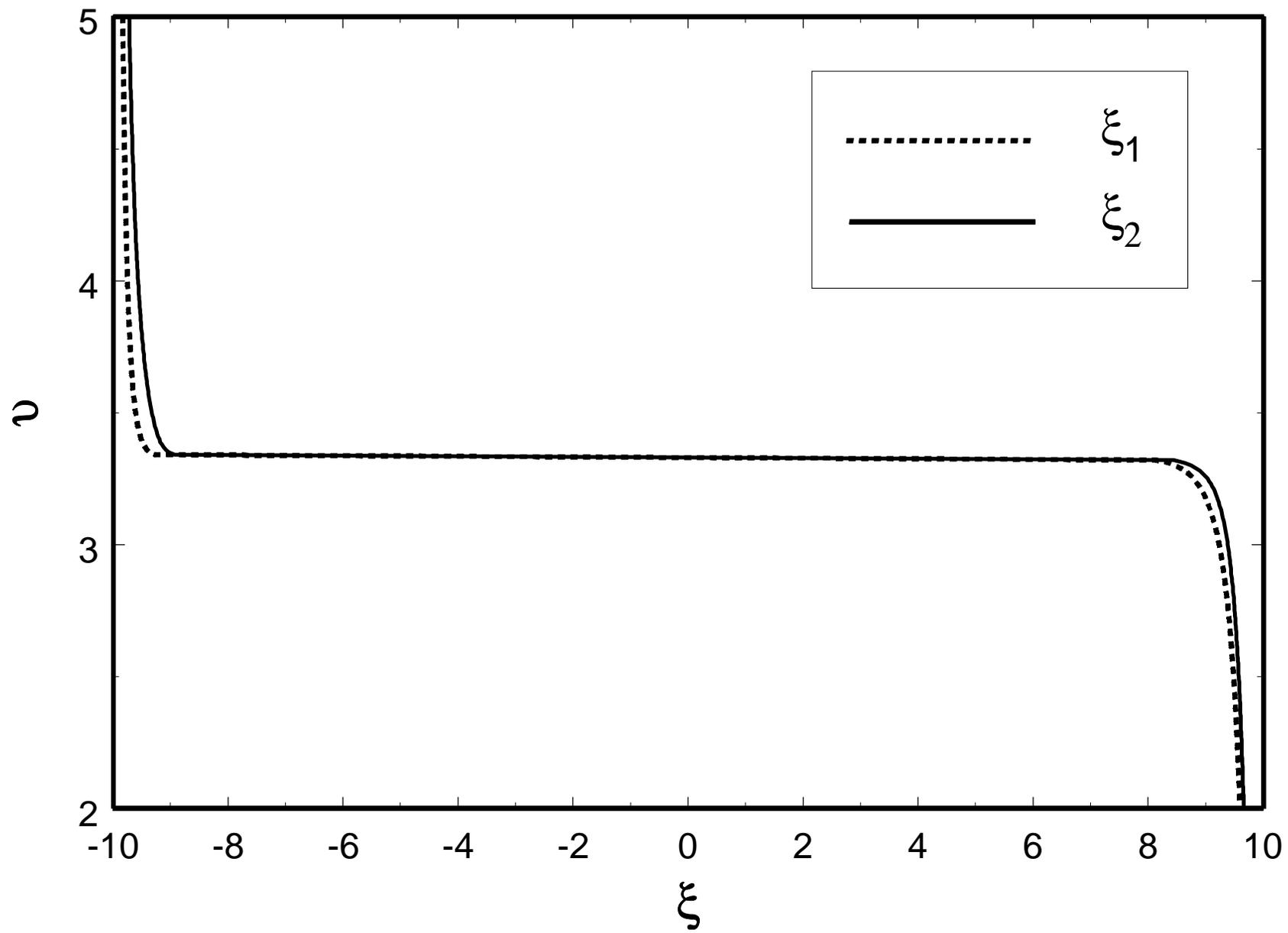
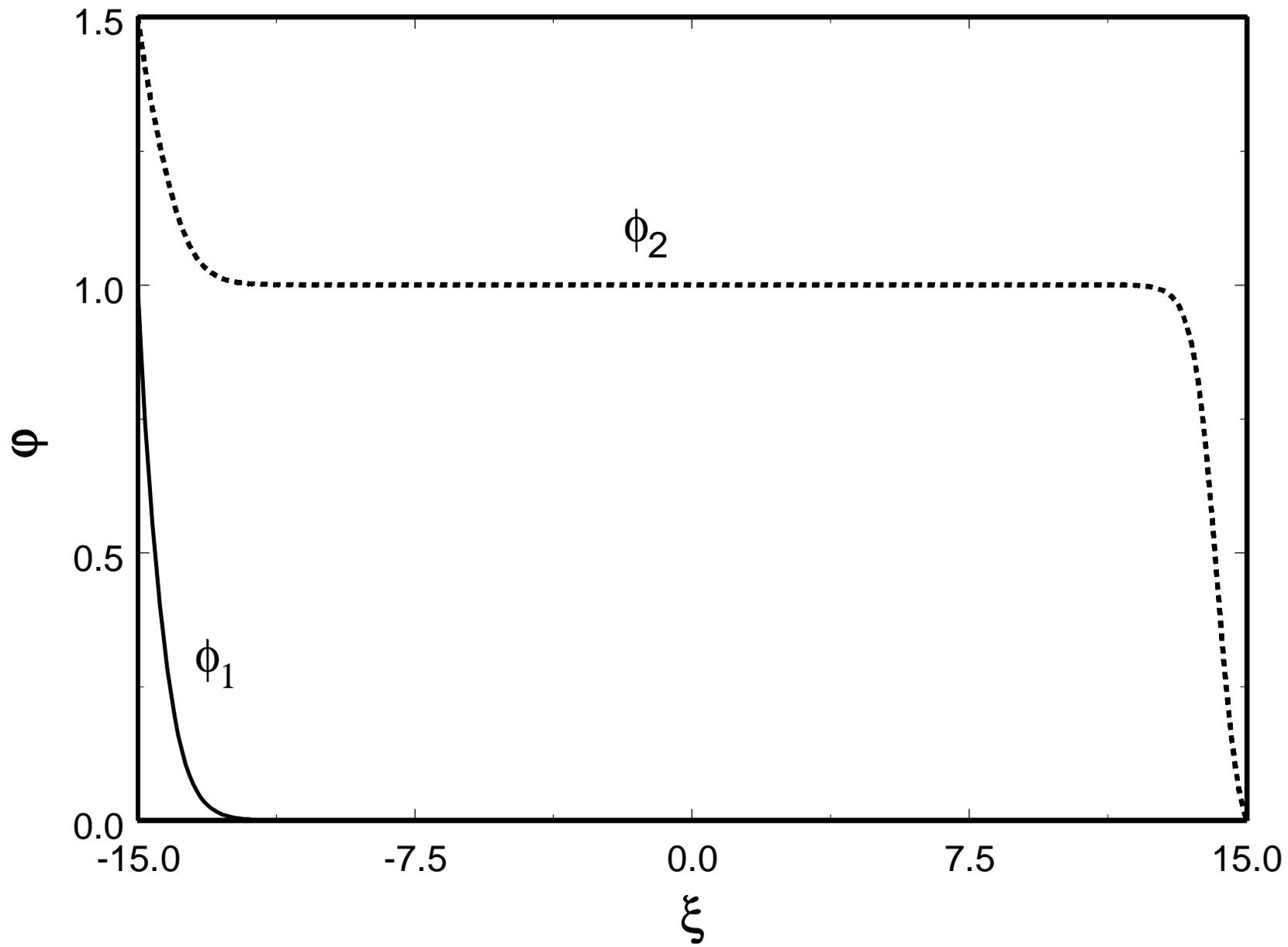
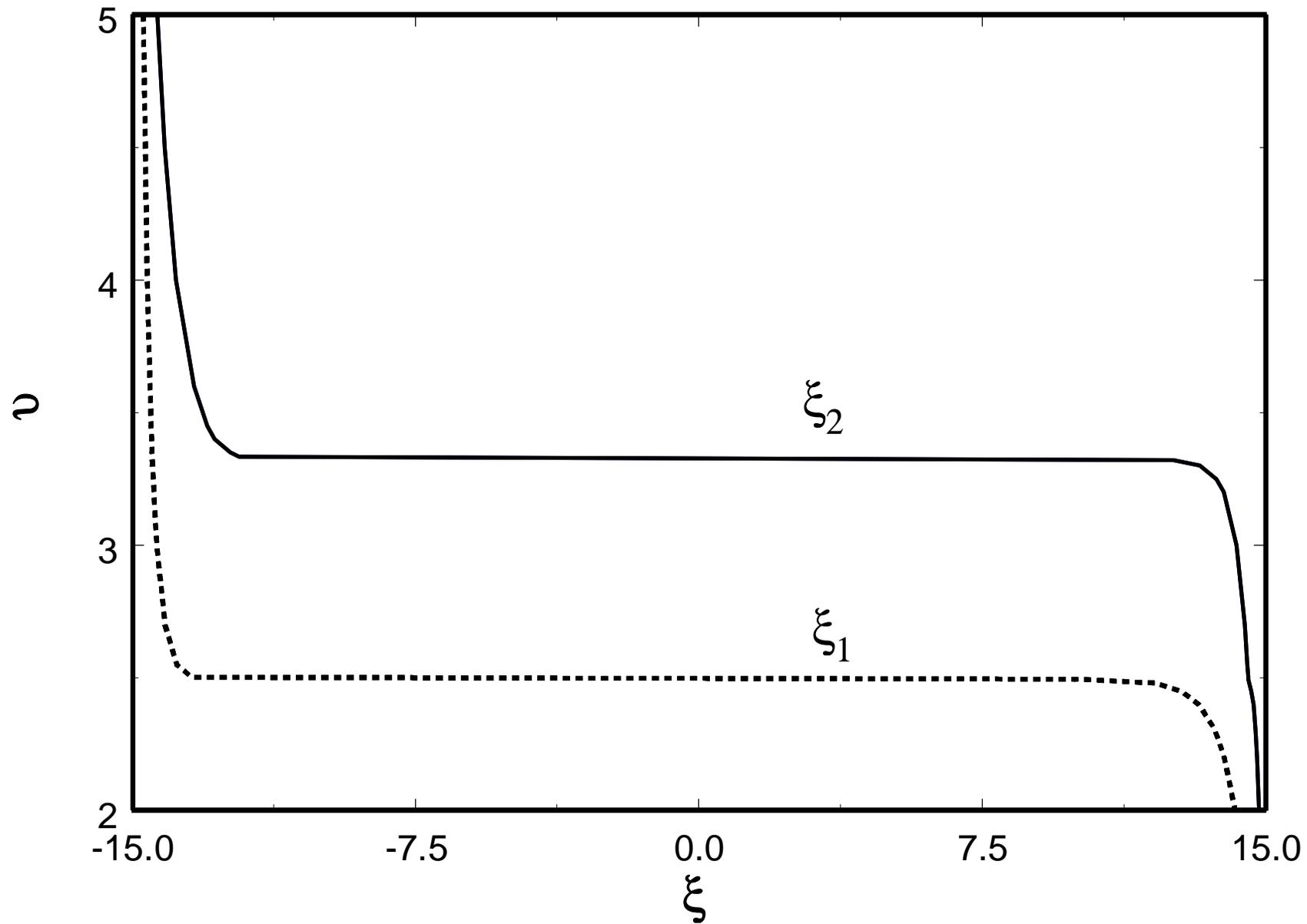
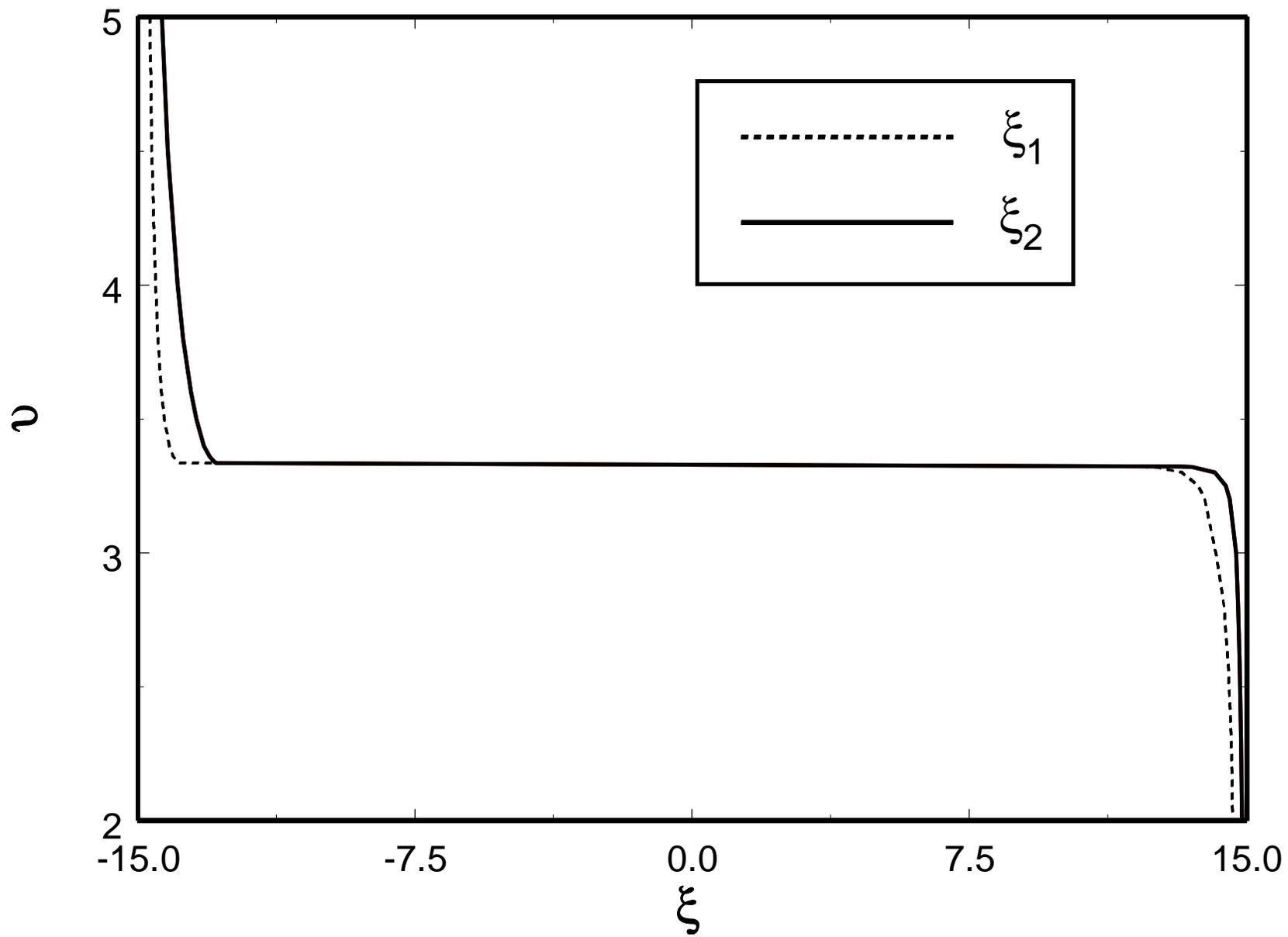
\end{document}